\documentclass[twocolumn]{aastex63}
\usepackage{comment}

\usepackage{amsmath}
\usepackage{mathrsfs} 
\usepackage{graphicx}
\usepackage{ulem}
\usepackage{soul}
\usepackage{xcolor}
\usepackage{macros_vdb}

\received{March 2, 2021}
\revised{May 11, 2021}
\accepted{May 22 2021}

\shorttitle{Random Motion of Nuclear Objects in FDM Halos}
\shortauthors{Dutta Chowdhury et al.}

\begin{document}

\title{On the Random Motion of Nuclear Objects in a Fuzzy Dark Matter Halo}

\correspondingauthor{Dhruba Dutta Chowdhury}
\email{dhruba.duttachowdhury@yale.edu}

\author[0000-0003-0250-3827]{Dhruba Dutta Chowdhury}
\affiliation{Department of Astronomy, Yale University, New Haven, CT-06511, USA} 

\author[0000-0003-3236-2068]{Frank C. van den Bosch}
\affiliation{Department of Astronomy, Yale University, New Haven, CT-06511, USA}

\author{Victor H. Robles}
\affiliation{Yale Center for Astronomy and Astrophysics, New Haven, CT-06520, USA}

\author[0000-0002-8282-9888]{Pieter van Dokkum}
\affiliation{Department of Astronomy, Yale University, New Haven, CT-06511, USA}

\author[0000-0002-1249-279X]{Hsi-Yu Schive}
\affiliation{Department of Physics, National Taiwan University, Taipei 10617, Taiwan}
\affiliation{Institute of Astrophysics, National Taiwan University, Taipei 10617, Taiwan}
\affiliation{Center for Theoretical Physics, National Taiwan University, Taipei 10617, Taiwan}
\affiliation{Physics Division, National Center for Theoretical Sciences, Taipei 10617, Taiwan}

\author{Tzihong Chiueh}
\affiliation{Department of Physics, National Taiwan University, Taipei 10617, Taiwan}
\affiliation{Institute of Astrophysics, National Taiwan University, Taipei 10617, Taiwan}
\affiliation{Center for Theoretical Physics, National Taiwan University, Taipei 10617, Taiwan}

\author{Tom Broadhurst}
\affiliation{Department of Theoretical Physics, University of the Basque Country UPV/EHU, E-48080 Bilbao, Spain}
\affiliation{Donostia International Physics Center (DIPC), 20018 Donostia-San Sebastian (Gipuzkoa), Spain}
\affiliation{Ikerbasque, Basque Foundation for Science, E-48011 Bilbao, Spain}

\begin{abstract}
 Fuzzy Dark Matter (FDM), consisting of ultralight bosons ($m_\rmb \sim 10^{-22} \eV$), is an intriguing alternative to Cold Dark Matter. Numerical simulations that solve the Schr\"odinger-Poisson (SP) equation show that FDM halos consist of a central solitonic core, which is the ground state of the SP equation, surrounded by an envelope of interfering excited states. These excited states also interfere with the soliton, causing it to oscillate and execute a confined random walk with respect to the halo center of mass. Using high-resolution numerical simulations of a $6.6 \times 10^9 \Msun$ FDM halo with $m_\rmb = 8 \times 10^{-23} \eV$ in isolation, we demonstrate that the wobbling, oscillating soliton gravitationally perturbs nuclear objects, such as supermassive black holes or dense star clusters, causing them to diffuse outwards. In particular, we show that, on average, objects with mass $\lta 0.3 \%$ of the soliton mass ($M_{\rm sol}$) are expelled from the soliton in $\sim 3\Gyr$, after which they continue their outward diffusion due to gravitational interactions with the soliton and the halo granules. More massive objects ($\gtrsim 1 \% M_{\rm sol}$), while executing a random walk, remain largely confined to the soliton due to dynamical friction. We also present an effective treatment of the diffusion, based on kinetic theory, that accurately reproduces the outward motion of low mass objects and briefly discuss how the observed displacements of star clusters and active galactic nuclei from the centers of their host galaxies can be used to constrain FDM.
\end{abstract}


\section{Introduction}\label{sec:Intro}

In the standard Lambda Cold Dark Matter ($\Lambda$CDM) model of cosmic structure formation, dark matter is assumed to be collisionless and to consist of weakly interacting, massive particles (WIMPs) with a rest mass energy of order a few GeV. While extremely successful on large scales, it faces a barrage of `small scale problems', such as the cusp-core problem \citep[][]{moore94, floresetal94, deblok10, ohetal11}, the missing satellites problem \citep[e.g.,][]{klypin99, mooreetal99}, and the too-big-to-fail problem \citep[e.g.,][]{boylan-kolchin11, boylan-kolchin12, tollerud14}. Although properly accounting for the impact of baryons can largely alleviate these problems \citep[see][for a review]{bullock17}, their persistence in the literature, together with the dearth of any direct evidence for either super-symmetry or WIMPs \citep[e.g.,][]{liu17}, has resulted in a surge of alternative dark matter models.  

One such alternative is fuzzy dark matter (FDM), also known as scalar field dark matter or wave dark matter, which postulates that dark matter consists of ultra-light bosonic particles with masses $m_\rmb \sim 10^{-22} \eV$ (e.g., \citealp{guzmanetal00, peebles00, hu00}; see review papers by \citealp{suarezetal14}, \citealp{hui17}, \citealp{niemeyer20}, and \citealp{hui21}). An example of such bosons are axion-like particles that naturally arise in string theory \citep[e.g.,][]{svrcek06,arvanitaki10,luu20}. The typical de-Broglie wavelength, $\lambda_{\rm db} = h /(m_\rmb\,\sigma)$, where $h$ is the Planck's constant, for such particles is of the order of a few kpc in halos with a velocity dispersion, $\sigma$, of the order of $100 \kms$. On scales $\lambda \lta \lambda_{\rm db}$, the uncertainty principle gives rise to a large quantum pressure, which counters gravitational collapse. Consequently, structure formation in FDM models is suppressed on small scales, which provides a possible solution to the missing satellites problem \citep[e.g.,][]{roblesetal15, schive16, kulkarni20, may21}. On scales much larger than the de Broglie wavelength, FDM behaves just like CDM.

Since the occupation numbers of the bosonic FDM density field are huge, FDM behaves as a classical field, characterized by a wavefunction, $\psi$, that obeys the Schr\"odinger equation for a self-gravitating particle in a potential that relates to the density, $\rho = m_\rmb |\psi|^2$, via the Poisson equation. Numerical simulations \citep[e.g.,][]{schive14a, schwabe16, mocz17, veltmaat18} show that FDM halos consist of a constant density core, which helps to alleviate the cusp-core and too-big-to-fail problems \citep[e.g.,][]{roblesetal19}, surrounded by an envelope with a density profile that is similar to the NFW profile \citep[][]{navarro97} of CDM halos. The central core, also known as the soliton, is the ground-state of the Schr\"odinger-Poisson (SP) equation. The envelope consists of excited states of the SP equation, which extensively interfere with one another, giving rise to density fluctuations (`wave-granularity') that are correlated on scales comparable to the de Broglie wavelength.

If the halo were to remain in isolation, these fluctuations would weaken over time as the system expels probability to infinity, a process known as `gravitational cooling' \citep[e.g.,][]{guzman06}, such that the halo ultimately relaxes towards a naked soliton core. Given, though, that the associated relaxation time is typically larger than or of the order of the Hubble time \citep[][]{hui17} and that halos are continuously perturbed due to ongoing mass accretion and/or encounters with other halos, a typical FDM halo is characterized by order unity density fluctuations, which have various observable effects. In particular, the fluctuations act as a heating source, exerting random kicks to particles that move within it. This can lead to the thickening of stellar disks \citep[][]{church19, elzant20a} and streams \citep[][]{amorisco18, dalal20}. The heating effect of FDM can also counter dynamical friction, thereby causing the inspiral of black holes or star clusters to stall before they reach the halo center \citep{bar-or19}. 

A soliton in isolation, being an eigenstate of the SP equation, is expected to have a spherically symmetric, time-invariant density profile\footnote{Being a stationary state, time evolution only affects the phase of the soliton's wave function}. However, simulations show that the soliton core of an FDM halo exhibits strong temporal oscillations \citep[][]{veltmaat18}, while simultaneously executing a confined random walk within the central region of the halo, with a characteristic displacement of order its own size \citep[][]{schive20, li20}. As discussed in \citet{li20}, these arise from interference between the ground state (the soliton) and the excited states that make up the halo surrounding the soliton. An alternative, more classical, way of viewing this is that the soliton is perturbed by the order unity density fluctuations, which can be treated as short-lived quasiparticles \citep[][]{hui17}, in the halo that surrounds it. 

These oscillations and random walk of the soliton can have important implications for baryonic objects. For example, \citet{marsh19} discuss how the temporal oscillations of the soliton can heat (i.e., puff-up) the nuclear star cluster in Eridanus-II (but see also \citealp{chiang21} who argue that soliton oscillations cannot efficiently heat the Eridanus-II star cluster, as the oscillation time period is much larger than orbital timescales within the cluster). An additional phenomenon, not considered by \citet{marsh19} or \citet{chiang21}, is the center of mass motion of a nuclear star cluster, initially at rest at the soliton center, with respect to the soliton. While the soliton is subject to both gravity and gradients in quantum pressure from the halo envelope, the cluster only feels the former. Hence, the two will respond differently, initiating a random walk of the cluster with respect to the soliton, in which the cluster's orbit continues to be gravitationally perturbed by the wobbling, oscillating soliton. As shown in \citet{schive20}, the wobble of the soliton also contributes significantly to the heating of star clusters and results in tidal distortions that can completely disrupt low-density star clusters such as the one in Eridanus II in $\sim 1 \Gyr$. However, if the cluster is dense enough, it can be expelled from the soliton with minimal tidal disruption, after which potential fluctuations in the halo envelope (arising both from the soliton and the quasiparticles) will take over and continue to `push' the cluster outwards until it is in equilibrium with the halo. The same mechanism will also operate on supermassive black holes (SMBHs) located within the confines of the soliton.

Hence, statistics on the observed displacements of nuclear star clusters \citep[e.g.,][]{binggeli00, cote06} and active galactic nuclei \citep[AGN, e.g.,][]{shen19, reines20} from the centers of mass of their host galaxies can be used to constrain FDM. We note that the outward diffusion of SMBHs in FDM halos has previously been discussed in \citet[][]{elzant20b} using analytic prescriptions. However, they only account for the diffusion due to the quasiparticles outside of the soliton and assume that some external mechanism (i.e., a galaxy-galaxy merger) is required to initially eject the SMBH from the soliton. The study presented here will show that even in the absence of external causes, on average, SMBHs diffuse out of the soliton on a relatively short timescale.

In this paper, we use self-consistent, high-resolution numerical simulations solving the SP equation for an FDM halo of mass $\sim 6.6 \times 10^{9} M_{\odot}$ (extracted from a cosmological FDM simulation with $m_\rmb=8 \times 10^{-23} \eV$), in which we inject point particles of masses in the range  $10^{5}-10^{7}\ M_{\odot}$, representing SMBHs or dense star clusters. The particles are initially placed at rest at the center of the soliton, and their subsequent evolution is tracked for $20 \Gyr$. Using statistical samples, we investigate how objects of different mass are dispersed away from the center.

Note that there are several existing constraints on the FDM particle mass, including those that derive from the Lyman-alpha forest: $m_\rmb \gtrsim 3 \times 10^{-21} \eV$ \citep[e.g.,][]{armengaud17,irsic17,kobayashi17,nori19}, the abundance of Milky Way (MW) subhalos: $m_\rmb \gtrsim 2-5 \times 10^{-21} \eV$ \citep[e.g.,][]{nadler19,benito20,schutz20,descollab21}, and the galaxy luminosity function at high redshifts: $m_\rmb \gtrsim 1-8 \times 10^{-22} \eV$ \citep[e.g.,][]{schive16,corasaniti17,menci17}. In addition, CMB data constrain the relic density of ultra-light bosons to be within $5 \%$ of the total dark matter relic density for $10^{-32} \eV \leq m_\rmb \leq 10^{-25.5} \eV$. For $m_\rmb \gtrsim 10^{-24} \eV$, though, ultra-light bosons are indistinguishable from CDM on the length scales probed by the CMB data and are, therefore, allowed \citep[][]{hlozek15}. Constraints derived from comparing dynamical data of MW dwarfs with FDM predictions yield the following results: while a larger $m_\rmb$ ($\sim 3.7-5.6 \times 10^{-22} \eV$) is required to fit the data for ultra-faint dwarfs such as Draco-II and Triangulum-II \citep[][]{calabrese16}, a smaller $m_\rmb$ ($\lesssim 0.4-1.1 \times 10^{-22} \eV$) is favoured by classical dwarfs such as Sculptor and Fornax (\citealp{marsh15}, \citealp{gonzalez-moralez17}, see also \citealp{safarzadeh20}). Some other constraints include $m_\rmb > 0.6 \times 10^{-22} \eV$ to avoid overheating of the MW disk \citep[][]{church19} and  $m_\rmb > 1.5 \times 10^{-22} \eV$ inferred from MW globular cluster stellar streams \citep[][]{amorisco18}. While some of these constraints are more stringent and seem to exclude a $\sim 10^{-22} \eV$ ultra-light boson, we emphasize that each of them  are subject to numerous sources of systematic uncertainties. Consequently, our assumed boson mass of $m_\rmb=8 \times 10^{-23} \eV$ can not be ruled out with certainty.

\begin{figure*}
\centering
\includegraphics[width=1.\textwidth]{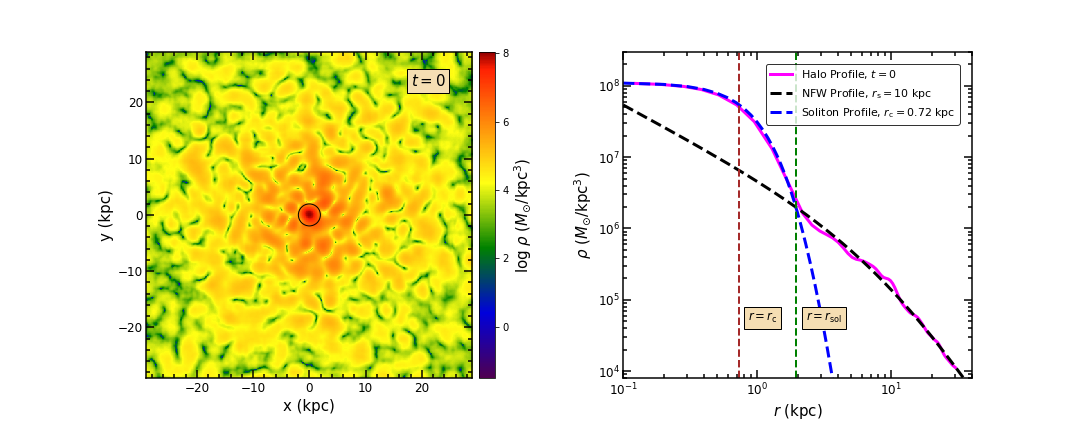}
\caption{The left-hand panel shows the density of our FDM halo at $t=0$ in a $x$-$y$ slice that cuts through the maximally dense cell. The halo consists of a ground state, also known as the soliton (central, red nugget), which is surrounded by a much larger region of excited states that interfere with one another and give rise to spatial fluctuations in density. The magenta curve in the right-hand panel shows the density profile of the halo for this snapshot, averaged over shells of radius $r$ to $r+\Delta r$, where $r=0$ is the location of the maximally dense cell. It is soliton-like (dashed, blue curve) at small radii, out to about $r_{\rm sol}=2.7 r_{\rm c}$ (dashed, green, vertical line), where $r_{\rm c}=0.72\ \rm kpc$ is the core radius of the soliton (dashed, brown, vertical line) and NFW-like (dashed, black curve) at large radii. $r_{\rm sol}$ is assumed to be the soliton boundary, roughly indicating the radius where the soliton ends, and the rest of the halo begins. The black circle marks this boundary in the left-hand panel.}
\label{fig:halo_initial}
\end{figure*}

The paper is organized as follows. In Section~\ref{sec:isolated_fdm_halo}, we present a detailed discussion on the evolution of the FDM halo in isolation. Section~\ref{sec:particles} describes the results of the simulations in which we inject massive point particles at rest at the center of the soliton. In Section~\ref{sec:discusssion}, we compare these results with theoretical predictions based on kinetic theory. We summarize and conclude in Section~\ref{sec:concl}.

\section{FDM Halo in Isolation}
\label{sec:isolated_fdm_halo}

Throughout this paper, we simulate a single FDM halo with a virial mass of $\Mvir = 6.6 \times 10^{9} \Msun$, extracted from a large, cosmological simulation of structure formation in a universe with a bosonic dark matter particle mass of $m_\rmb = 8 \times 10^{-23} \eV$ \citep[see][for details]{schive14a}. We extract the wave function from a region of $62.5\kpc \times 62.5\kpc \times 62.5\kpc$ centered on this halo from the redshift zero output, which we then simulate at a uniform spatial resolution of $\Delta x = 122\pc$ using the code {\tt GAMER-2} \citep{schive18}, which evolves the system by solving the SP equation. We adopt periodic boundary conditions when updating the wave function (i.e., flow traveling across the right edge will re-enter the simulation domain from the left edge) and isolated boundary conditions (i.e., the potential is zero at infinity) when computing the gravitational potential. As discussed later, the soliton radius is about $2\ \rm kpc$ (or about $16\ \Delta x$), and the halo density fluctuations are roughly soliton-sized. Therefore, at this spatial resolution, both the soliton and the rest of the halo are well resolved. Throughout, we adopt a time step of $\Delta t=1.57 \times 10^{5} \yr$, as inferred from the stability consideration of the kinetic and potential energy operators in the SP equation \citep[see][for details]{schive14a}. The total halo mass within the simulation box is $5.7 \times 10^{9} M_{\odot}$, which is about $86 \%$ of $M_{\rm vir}$. 

We let the isolated halo evolve for several $\Gyr$ in order to attain equilibrium. In what follows, we take this evolved, equilibrium state as our initial conditions, corresponding to $t=0$. Next, we evolve the halo for an additional $10 \Gyr$ to study its properties in isolation. During this period, the mass and total energy of the halo are conserved to better than $0.0003 \%$ and $0.6 \%$, respectively. 

\subsection{Density Profile} 
\label{sec:density_profile}

Prior simulations \citep[e.g.,][]{schive14a, schive14b, mocz17, veltmaat18} have shown that an FDM halo consists of a ground state, also known as the soliton, which is surrounded by a much larger region of excited states that interfere with one another. This general behavior is illustrated in Figure~\ref{fig:halo_initial}, the left-hand panel of which shows the density of our FDM halo at $t=0$ in a $x$-$y$ slice that cuts through the maximally dense cell. The central, red nugget indicates the soliton. Outside the soliton, spatial fluctuations in density are evident throughout the halo, resulting from the interference between the excited states. The magenta curve in the right-hand panel shows the density profile of the halo for this snapshot, averaged over shells of radius $r$ to $r + \Delta r$, where $r=0$ is the location of the maximally dense cell. 

As shown in \citet[][]{schive14b}, solitonic cores have a universal density profile that is well fit by
\begin{equation}
    \rho_{\rm sol}(r) = \frac{\rho_0}{[1+0.091\ (r/r_\rmc)^{2}]^8} \ .
    \label{sol_prof}
\end{equation}
Here, $\rho_0$ is the central density and $r_\rmc$ is the core radius, defined such that $\rho_{\rm sol}(r_\rmc) = \rho_{0}/2$. In addition, because of the scaling symmetry of the SP equation \citep[see e.g.,][]{seidel90,guzman06}, the core radii of solitonic cores obey the following scaling relation
\begin{equation}
\left( \frac{r_\rmc}{\kpc} \right)^4 = \left( \frac{1.954 \times 10^{9} \Msun \kpc^{-3}}{\rho_{0}} \right) \left( \frac{m_\rmb}{10^{-23} \eV} \right)^{-2} \ .
\label{core_radius}
\end{equation} 
The dashed, blue curve in Figure~\ref{fig:halo_initial} indicates the universal solitonic density profile of Equation~\ref{sol_prof}, in which we adjust $\rho_0$ to the value of the densest cell in our simulation, while $r_\rmc$ is obtained from the above scaling relation, which yields $r_\rmc = 0.72 \kpc$, indicated by the brown, dashed, vertical line. The corresponding mass of the soliton is $M_{\rm sol} \simeq 11.59 \rho_0 r^3_\rmc \simeq 4.5 \times 10^8\Msun$. Note that the resulting density profile is an excellent fit to that of our solitonic core. 

The dashed, black curve is an NFW profile given by
\begin{equation}
\rho_{\rm NFW}=\frac{\rho_{\rm s}}{(r/r_{\rm s}) (1+r/r_{\rm s})^2} \ ,
\end{equation}
with $r_{\rm s}=10\ \rm kpc$ and $\rho_{\rm s}=5.5 \times 10^{5} M_{\odot} \rm kpc^{-3}$, chosen such that it matches the slope and normalization of the density profile in the outer halo. The dashed, green, vertical line at $r_{\rm sol}=2.7 r_{\rm c}$ roughly indicates the radius where the halo's density profile begins to deviate from the soliton profile, becoming NFW-like at larger radii. In what follows, we consider $r_{\rm sol}$ to indicate the boundary of the soliton. The black circle in the left-hand panel of Figure~\ref{fig:halo_initial} marks this boundary.

\begin{figure*}
  \centering
  \includegraphics[width=0.9\textwidth]{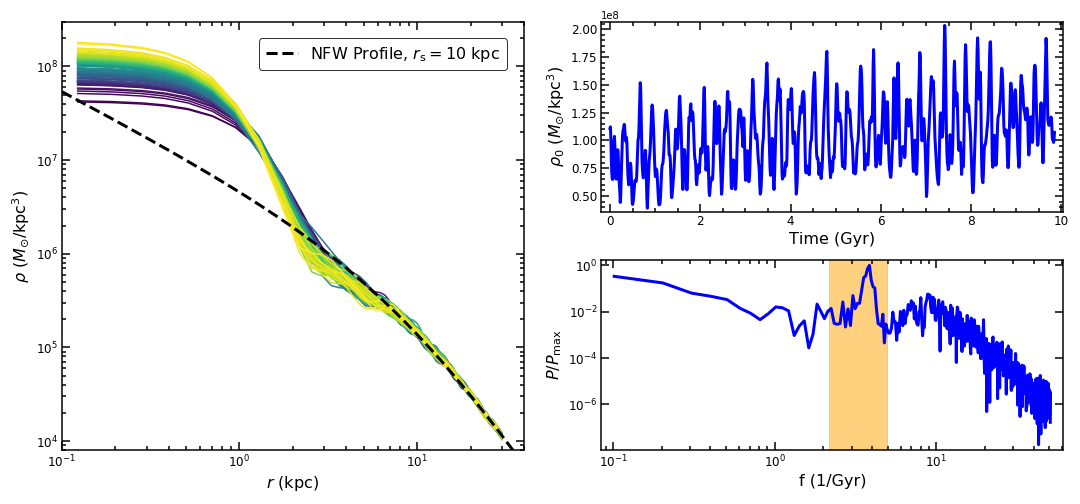}
  \caption{The different curves in the left-hand panel show the shell-averaged density profile of the halo at 100 equally spaced time instants between $0$ to $10\Gyr$, centered on the maximally dense cell at each time. They are color-coded by the central density, $\rho_{0}$, varying from yellow to violet with decreasing $\rho_{0}$. The soliton undergoes temporal oscillations, perturbed by the density fluctuations in its immediate surroundings. When $\rho_{0}$ increases (decreases), the soliton shrinks (expands) in size, in accordance with Equation~\ref{core_radius}. The halo region close to the soliton (and interacting with it) also experiences a decrease (increase) in density, which follows from the conservation of the total halo mass. The shell-averaged density in the halo outskirts remains NFW-like (black, dashed line) all throughout. The upper and lower right-hand panels show the oscillations in $\rho_{0}$ and the corresponding Fourier power spectrum, respectively, using 1000 outputs between $0$ to $10\Gyr$. The orange, shaded region in the lower right-hand panel highlights the characteristic frequency of these oscillations predicted by \citet[][]{veltmaat18}, which is in good agreement with the frequency range over which the power spectrum peaks in our simulation (see text for details).}
\label{fig:density_time_evolve}
\end{figure*}

The different curves in the left-hand panel of Figure~\ref{fig:density_time_evolve} show the shell-averaged density profile of the halo at 100 equally spaced instants between $t=0$ and $10 \Gyr$. Similar to Figure~\ref{fig:halo_initial}, $r=0$ is the location of the maximally dense cell. These curves are color-coded by the central density, varying from yellow to violet with decreasing $\rho_{0}$. Although the halo is in virial equilibrium, the soliton is undergoing order unity density fluctuations. As discussed in \citet[][]{li20}, these arise from interference between the ground state and excited states, or equivalently, from interactions between the soliton and the surrounding density fluctuations \citep[see also][]{veltmaat18}. Note how the soliton shrinks in size when its central density is boosted, and vice versa, in accordance with Equation~\ref{core_radius}. Additionally, as the total halo mass is conserved, when the soliton becomes more (less) concentrated, the region of the halo close to the soliton (and interacting with it) experiences a decrease (increase) in density.

In the upper and lower right-hand panels, we show the oscillations in the central density and the corresponding Fourier power spectrum, respectively, using 1000 outputs between $0$ to $10\Gyr$, corresponding to a time resolution of $0.01\ \rm Gyr$. The central density fluctuates by about a factor of $2$ around a time-averaged value of $1.04 \times 10^{8} M_{\odot} \rm kpc^{-3}$. The characteristic frequency of these oscillations is given by 
\begin{equation}
f = 10.94 \Gyr^{-1} \, \left( \frac{\rho_{0}}{10^{9} M_{\odot} \rm kpc^{-3}} \right)^{1/2}\, ,
\label{oscillation_freq}
\end{equation}
\citep[see][]{veltmaat18}. Using the maximum and minimum values of $\rho_{0}$ from the time-series in the upper right-hand panel, we obtain $f_{1}=4.93\ \rm Gyr^{-1}$ and $f_{2}=2.16\ \rm Gyr^{-1}$, respectively, corresponding to periods of $0.2-0.5 \Gyr$. The region bounding $f_1$ and $f_2$ is shaded in orange in the lower right-hand panel and is in good agreement with the frequency range over which the power spectrum peaks in our simulation. Note that the core radius of the soliton, which can be determined using Equation~\ref{core_radius}, also oscillates with the same characteristic frequency by a factor of $\sim 1.2$ around a time-averaged value of $0.75\ \rm kpc$. Therefore, the time-averaged value for $r_{\rm sol}$ is $2\ \rm kpc$. 

\subsection{Velocity Dispersion Profile} 
\label{sec:velocity_dispersion_profile}

Besides density, another important physical property of a dark matter halo is its velocity dispersion, which quantifies the amount of random motion within the halo. The velocity dispersion of our FDM halo can be calculated from the velocities of the simulated cells. The velocity of each cell is given by ${\bf v}=(\hbar/m_{\rmb}) \bf{\nabla} \theta$, where $\hbar$ is the reduced Planck constant and $\theta$ is the phase of the wavefunction. For each snapshot, the velocity dispersion, $\sigma_{h}(r)$, for a shell of radius $r$ to $r+\Delta r$ is then defined as

\begin{equation}
\sigma_{\rm h}(r)=\sqrt{\frac{\sigma_{x}^2+\sigma_{y}^2+\sigma_{z}^2}{3}} \, ,
\label{dispersion}
\end{equation}
where
\begin{equation}
\sigma_{i}^2= \frac{\langle \rho v^2_{i} \rangle}{\langle \rho \rangle}-\frac{\langle \rho v_{i} \rangle^2}{\langle \rho \rangle^2} \, , \hspace{0.2cm} i=x,y,z \, .
\label{dispersion_1}
\end{equation}
Here, $v_{i}$ is the velocity of a cell in the $i^{\rm th}$ direction, $\rho$ is its density, and the angle brackets denote averaging over the shell. Therefore, $\sigma_{i}^2$ is the mass-weighted variance in velocity along the $i^{\rm th}$ direction, which is obtained for the three independent directions and then averaged in quadrature to obtain the halo velocity dispersion as a function of $r$. The mass weighting is required because of the density fluctuations (cell-to-cell mass variance) within a shell.

\begin{figure}
  \centering
  \includegraphics[width=0.45\textwidth]{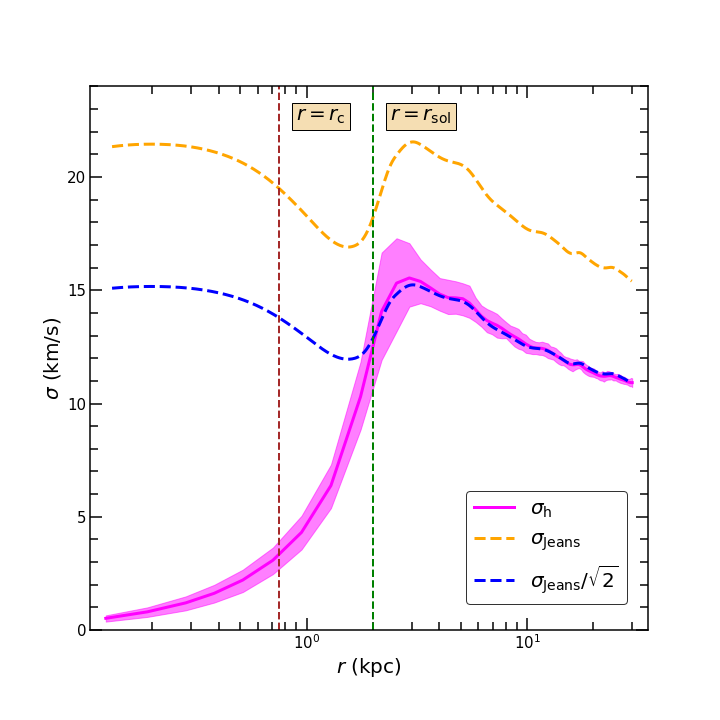}
  \caption{The magenta curve shows the time-and-shell averaged velocity dispersion, $\sigma_\rmh$, as function of radius, $r$, for our FDM halo. The associated envelope indicates the $16^{\rm th}-84^{\rm th}$ percentile range of shell-averaged values obtained at different times. The dashed, vertical, brown and green lines indicate the time-averaged core radius, $r_\rmc$ and soliton radius, $r_{\rm sol}$, respectively. For comparison, the dashed, orange curve plots the velocity dispersion profile, $\sigma_{\rm Jeans}$, obtained by solving the isotropic, spherical Jeans equation with the time-averaged density profile of the halo as input. The dashed, blue curve indicates $\sigma_{\rm Jeans}/\sqrt{2}$ and is in excellent agreement with the magenta curve outside $r_{\rm sol}$, indicating that the energy density outside the soliton has about equal contributions from kinetic energy (due to random motion) and quantum energy (due to quantum pressure).}
\label{fig:halo_dispersion}
\end{figure}

In Figure~\ref{fig:halo_dispersion}, the solid, magenta curve shows $\sigma_{\rm h}$ as a function of $r$, where $r=0$ is the maximally dense cell, averaged over 100 equally spaced snapshots between $0$ to $10 \Gyr$. The associated envelope highlights the $1 \sigma$ variation ($16^{\rm th}-84^{\rm th}$ percentile) over this period. The dashed, orange curve plots the velocity dispersion profile, $\sigma_{\rm Jeans}$, obtained by solving the spherical Jeans equation (under the assumption of velocity isotropy) with the time-averaged density profile of the halo (average over the different curves shown in the left-hand panel of Figure~\ref{fig:density_time_evolve}) as the input density. The dashed, vertical, brown and green lines indicate the time-averaged core radius, $r_\rmc$, and soliton radius, $r_{\rm sol}$, respectively.

As the self-gravity of an FDM halo is balanced by both random motion and quantum pressure \citep[see, e.g.,][]{hui17}, solving the Jeans equation gives an ``effective" dispersion, which is higher than the actual dispersion due to random motion. Using idealized soliton-soliton merger simulations, \citet{mocz17} have shown that in the final virialized halo, the contributions to the energy density at a particular radius from classical kinetic energy (due to random motion) and quantum energy (due to quantum pressure) are roughly equal outside the soliton. As the classical kinetic energy at a particular radius is proportional to $\sigma^2_{\rm h}$, this equipartition of energy suggests that outside the soliton, $\sigma_{\rm h}$ should be approximately equal to $\sigma_{\rm Jeans}/\sqrt{2}$. The blue, dashed curve plots $\sigma_{\rm Jeans}/\sqrt{2}$, which overlaps nicely with the magenta curve outside of the soliton ($r > 2\kpc$), as expected. Inside the soliton, $\sigma_{h}$ decreases with decreasing $r$, indicating decreasing random motion as $r \to 0$. Note that a naked soliton (without an associated halo) has the same phase throughout. Therefore, it has zero dispersion everywhere and is entirely supported by quantum pressure against collapse due to self-gravity. However, in our case, the soliton interacts with the rest of the halo, resulting in a non-zero dispersion inside $r_{\rm sol}$, which only tends to zero as $r \to 0$. 

\begin{figure*}
  \centering
  \includegraphics[width=1.\textwidth]{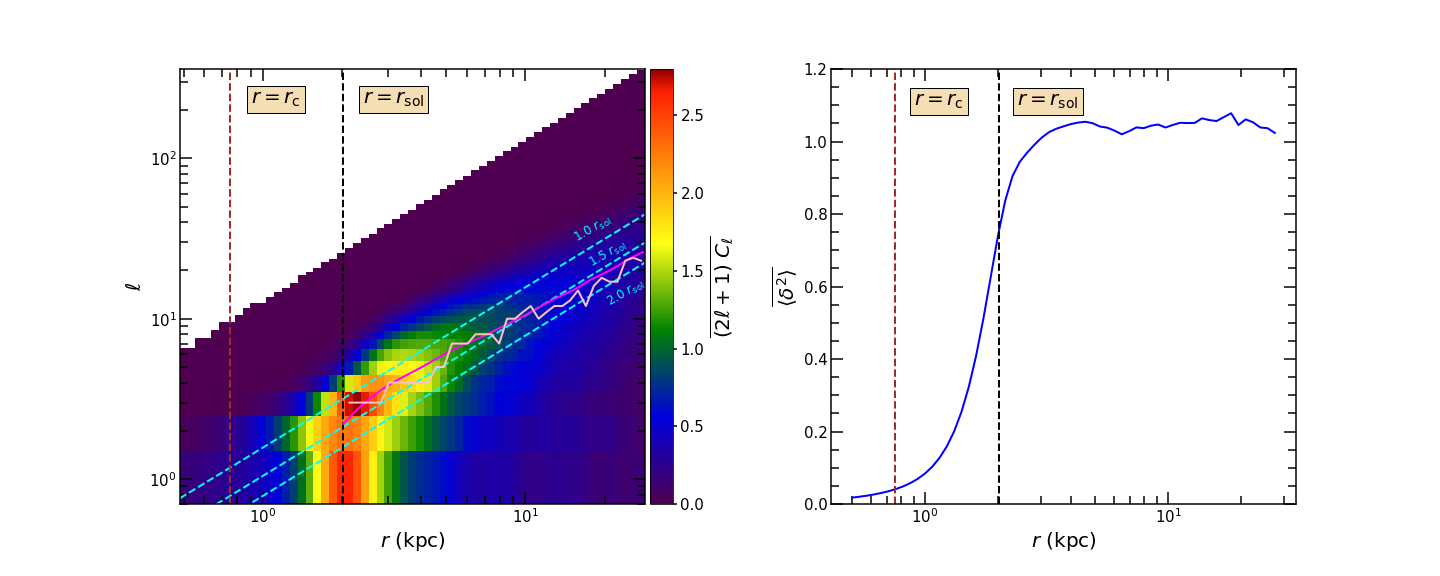}
  \caption{The left-hand panel shows the time-averaged multipole components, $\overline{(2l+1)C_\ell}$, of the variance in the overdensity field, $\delta$, as a function of radius, $r$, and angular wavenumber, $l$, where $r=0$ is the location of the maximally dense cell. The right-hand panel shows the time-averaged variance in $\delta$, $\overline{\langle \delta^{2} \rangle}$, as a function of $r$. In both panels, the brown and black, dashed, vertical lines indicate the time-averaged core radius ($r_\rmc = 0.75\kpc$) and boundary of the soliton ($r_{\rm sol} =2\kpc$), respectively. The soliton core ($r < r_\rmc$) is close to spherically symmetric, as is evident from the absence of significant power at all $\ell$ and the fact that $\overline{\langle \delta^{2} \rangle} \simeq 0$. Beyond $r_{\rm c}$, the soliton is aspherical, with a lot of low $\ell$ power and a gradually increasing $\overline{\langle \delta^{2} \rangle}$, as $r$ increases from $r_\rmc$ to $r_{\rm sol}$. Beyond $r_{\rm sol}$, the pink curve indicates the $\ell$ where the power peaks at a particular $r$, while the cyan lines highlight the $r$-$\ell$ relation for fluctuations of fixed size ($r_{\rm sol}$, $1.5r_{\rm sol}$, and $2r_{\rm sol}$), as labeled. Note that the characteristic size of the halo density fluctuations increases by about a factor of $2$ as $r$ increases from $\sim 2\kpc$ to $\sim 30\kpc$, which closely follows that inferred from the quasiparticle formalism \citep{bar-or19, chavanis20,elzant20a} and is indicated by the magenta curve. See text for details.}
\label{fig:average_power}
\end{figure*}

\subsection{Density Fluctuations} 
\label{sec:density_fluctuations}

The density fluctuations in the halo at a radius $r$ (from the maximally dense cell) and time $t$ can be characterized by defining the overdensity at a location ($\theta$, $\phi$, $r$) with respect to the mean density at that time and radius $\langle \rho(\theta, \phi|r, t) \rangle$ as
\begin{equation}
    \delta(\theta, \phi | r, t)=\frac{\rho(\theta, \phi | r, t) - \langle \rho(\theta, \phi | r, t) \rangle }{\langle \rho(\theta, \phi | r, t) \rangle} \, .
    \label{overdensity}
\end{equation}
Here, $\theta$ and $\phi$ are the usual spherical coordinates, and the angle brackets indicate an average over the sphere at radius $r$ at time $t$. The spherical harmonics expansion of $\delta$ is given by 
\begin{equation}
    \delta(\theta, \phi | r,t)= \sum_{\ell=0}^{\infty} \sum_{m=-\ell}^{\ell} a_{\ell}^{m}(r,t) Y_{\ell}^{m} (\theta, \phi) \ ,
    \label{harmonic_expansion}
\end{equation}
where $Y_{\ell}^{m} (\theta, \phi)$ are the usual spherical harmonics. As $Y_{\ell}^{m} (\theta, \phi)$ are orthonormal, the variance of $\delta$ at given $r$ and $t$ is given by
\begin{equation}
    \langle \delta^{2} \rangle (r,t)= \frac{1}{4 \pi} \sum_{\ell=0}^{\infty} (2\ell+1)C_\ell(r,t) \ ,
    \label{multipole_components}
\end{equation}
where 
\begin{equation}
    C_\ell(r,t)=\frac{1}{2\ell+1}\sum_{m=-\ell}^{\ell} |a_\ell^m(r,t)|^2 \ .
    \label{C_l}
\end{equation}

The left-hand panel of Figure~\ref{fig:average_power} plots the time-averaged total power $\overline{(2\ell+1)C_{\ell}}$ at a radius $r$ as a function of the angular wavenumber $\ell$ using 100 equally spaced outputs between $0$ to $10\Gyr$. The values of $r$ are sampled between $0.5\kpc$ and $27.5\kpc$, with a uniform logarithmic spacing of $\Delta\log r=0.03$. For each $r$ in each snapshot, the publicly available code {\it healpy} is used to make a pixel map with coordinates ($\theta$, $\phi$) such that each pixel has an equal area, and the angular resolution of the map is at least $\Delta x/r$, $\Delta x$ being the spatial resolution (cell width) of our simulation. Numerical values for $\rho$ are assigned to each {\it healpy} pixel by interpolating the density field. The overdensity, $\delta$ is then determined, as defined in Equation~\ref{overdensity}, and $C_{\ell}(r,t)$ is calculated for $0 \leq \ell \leq \ell_{\rm max}$. Here $\ell_{\rm max}=\pi r/(2 \Delta x)$, which corresponds to a physical scale of twice the simulation cell width. The results from all 100 snapshots are averaged to obtain $\overline{(2\ell+1)C_{\ell}}$ as a function of $r$ and $\ell$. The right-hand panel shows the time-averaged variance of $\delta$, $\overline{\langle \delta^{2} \rangle}$, as a function of $r$, which is determined by summing over the contribution from all time-averaged multipole components at that $r$ and dividing by $4\pi$. This follows from Equation~\ref{multipole_components}, upon time averaging both sides.

\begin{figure*}
  \centering
  \includegraphics[width=0.9\textwidth]{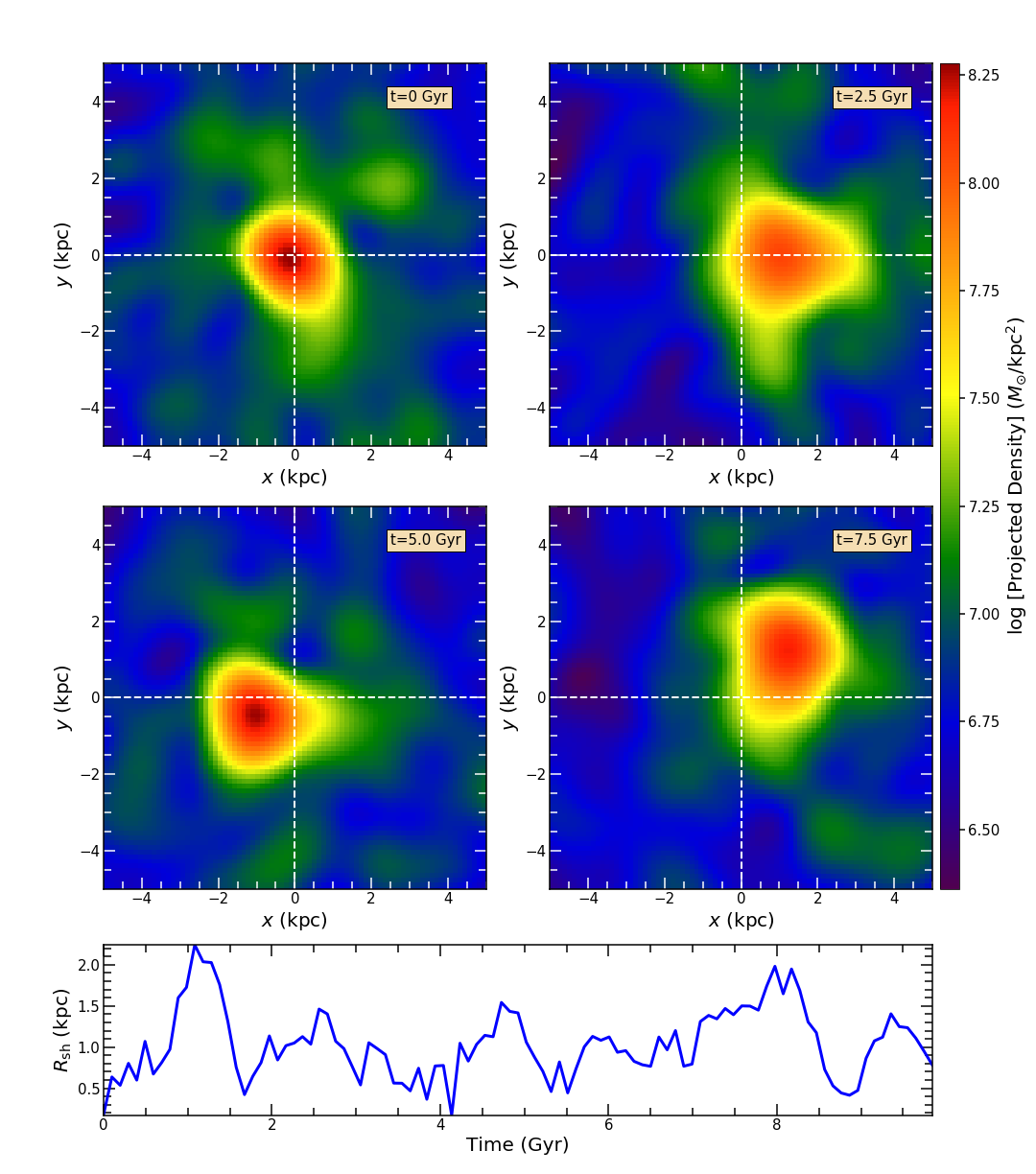}
  \caption{The four square panels show the projected density of our FDM halo in the $x$-$y$ plane at four different times, as labelled, in a region of $5\kpc \times 5\kpc$, centered on the center-of-mass of the halo. The soliton (red nugget) is clearly offset from the halo center-of-mass, and the offset changes with time. At any instant of time, there is a non zero net force acting on the soliton, exerted by the surrounding fluctuating density field, which causes this motion. In addition, the soliton experiences tidal forces that cause it to become aspherical in the outskirts. The bottom panel shows the offset of the maximally dense cell from the center-of-mass of the halo, $R_{\rm sh}$, as a function of time. The offset is less than or of the order of the soliton radius, its maximum value being $\sim 2.3\kpc$.}
\label{fig:random_walk}
\end{figure*}

From the left-hand panel, we note that there is very little power inside the core radius of the soliton, the brown, dashed, vertical line at $0.75\ \rm kpc$ indicating its time-averaged value. This is corroborated by the fact that $\overline{\langle \delta^{2} \rangle}$ is close to zero inside the core, as can be seen from the right-hand panel. Therefore, the soliton core is consistent with being spherically symmetric, without significant azimuthal fluctuations in the density. Between the core radius,  $r_{\rm c}$, and the soliton boundary, $r_{\rm sol}$,  $\overline{\langle \delta^{2} \rangle}$ increases from $0.05$ to $0.75$. As is evident from the left-hand panel, most of the associated power is concentrated at low $\ell$ ($\ell < 3$), indicating aspherical distortions of the soliton (see also Figure~\ref{fig:random_walk}). These are caused by the non-symmetric distribution of overdense and underdense regions around it, which exert tidal forces on the soliton. Moving outwards from the maximally dense cell, the soliton density decreases, and it becomes increasingly unstable against such perturbations.

The power spectrum outside of the soliton describes the density fluctuations in the rest of the halo. At each radius, $r$, the angular wavenumber corresponding to the maximum power, $\ell_{\rm p}$, is determined and overplotted on the power spectrum with the pink curve. From this curve, the characteristic size (diameter) of the fluctuations, $d$, can be inferred using $d=\pi r/\ell_{\rm p}$. The $\ell-r$ relationship for fixed characteristic sizes of $d=r_{\rm sol}$, $1.5 r_{\rm sol}$, and $2 r_{\rm sol}$ are indicated by the different cyan lines, as labeled. Comparing these lines with the pink curve shows that the characteristic size of the fluctuations increases roughly by a factor of two from $d \sim r_{\rm sol}$ just outside of the soliton to $d\sim 2 r_{\rm sol}$ at $r = 30 \kpc$. This is consistent with previous studies \citep{lin18,chan18} that inferred the typical size of the fluctuations from Fourier analysis of the overdensity field in radial shells.

\begin{figure*}
  \centering
  \includegraphics[width=0.9\textwidth]{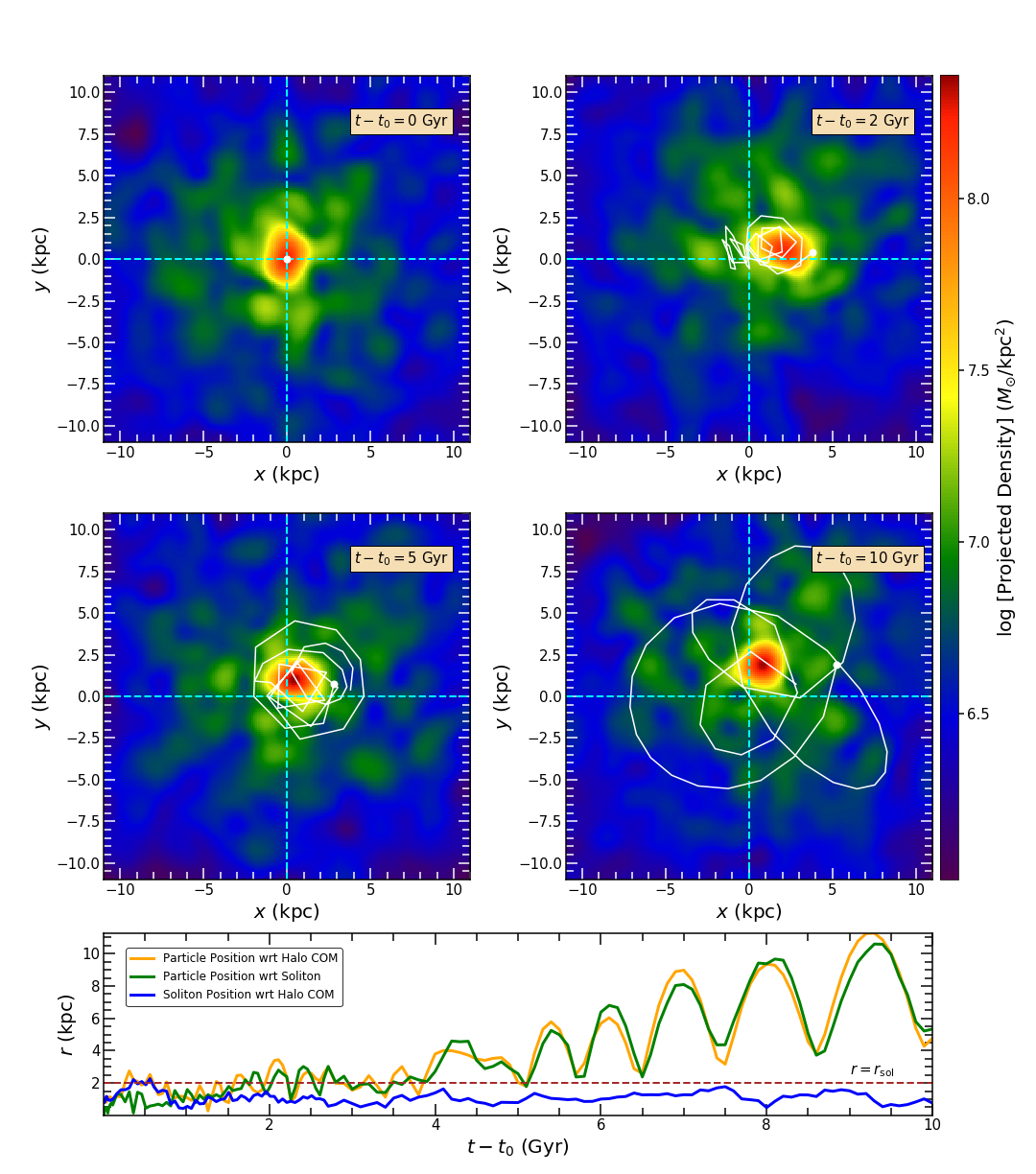}
  \caption{The four square panels from the top show the projected density of our FDM halo in the $x$-$y$ plane at four different times, as labeled, in a region of $10\kpc \times 10\kpc$, centered on the maximally dense cell at $t=t_{0}$. A $10^5 M_{\odot}$ particle is placed at the maximally dense cell (and at rest with respect to it) at $t=t_{0}$, indicated by the white circle in the top-left panel. The white curves in the top-left, middle-left, and middle-right panels show its subsequent orbital evolution between $0$-$2$, $2$-$5$, and $5$-$10$ Gyr respectively, the white circle in each case indicating the instantaneous position of the particle. It takes approximately $5\Gyr$ for the particle to diffuse out of the soliton. Once outside, it tries to settle into a rosette-like orbit, but continues to be gravitationally perturbed by interactions with the halo granules and the soliton. The bottom panel shows the orbital evolution of the particle with respect to the instantaneous soliton center (maximally dense cell) and the halo center of mass in green and orange, respectively. The motion of the soliton with respect to the halo center of mass is highlighted in blue.}
\label{fig:test_particle_motion}
\end{figure*}

As discussed in \citet{bar-or19}, \citet{chavanis20}, and \citet{elzant20a}, outside the soliton, the density fluctuations in an FDM halo can be envisioned as a sea of quasiparticles with an effective mass of
\begin{equation}
m_{\rm eff}=\rho (f \lambda_{\rm db})^{3} \, .
\label{effective_mass}
\end{equation}
Here, $\rho$ is the time-and-shell averaged density of the halo, 
\begin{equation}
    \lambda_{\rm db}=\frac{h}{m_{b} \sigma_{\rm Jeans}} \, ,
    \label{db}
\end{equation}
is the de Broglie wavelength, and $f$ is a constant that depends on the velocity distribution of the halo. Assuming the velocity distribution to be Gaussian, $f$ turns out to be $0.282$. The effective mass of the quasiparticles can be converted to an effective size by equating $m_{\rm eff}$ to $\rho V$, where V=$(\pi/6) d^3$ is the volume of a quasiparticle of diameter $d$. This yields $d \simeq 0.35 \lambda_{\rm db}$.\footnote{Note that it is also possible to define the de-Broglie wavelength in terms of the actual velocity dispersion of the halo, $\sigma_{\rm h}$, in which case $d \simeq 0.25\lambda_{\rm db}$.} 

Using this expression, we compute the local diameter of the quasiparticles as a function of $r$. The corresponding angular wavenumber, which is given by $\pi r/d$, is overplotted in the left-hand panel of Figure~\ref{fig:average_power} with the magenta curve. On comparing the pink and magenta curves, we find that the characteristic size of the halo density fluctuations outside the soliton inferred from our simulations is in good agreement with the effective size of the quasiparticles predicted from the theoretical models of \citet{bar-or19}, \citet{chavanis20}, and \citet{elzant20a}, thereby validating the quasiparticle picture. As to the variance in $\delta$ outside the soliton, it keeps increasing up to about $r=5 \kpc$, becoming roughly constant around a value of unity thereafter. 

\subsection{Soliton Random Walk} 
\label{sec:soliton_random_walk}

In addition to differential forces acting on the soliton that make it aspherical in the outskirts, the net force on the soliton from the surrounding fluctuating density field is also non-zero, which causes it to move as a whole with respect to the center of mass of the halo. In Figure~\ref{fig:random_walk}, we illustrate this motion by showing the projected density of the halo at four different time instants, as labeled, in a region of $5\ \rm kpc \times 5\ \rm kpc$, centered on the center of mass of the halo. The soliton (red nugget) is clearly offset from the halo center of mass, and the offset changes with time. The variation in the soliton density (the central density in particular) with time and the asphericity of the soliton are also evident from this figure, which concur with the results shown in Figures~\ref{fig:density_time_evolve} and \ref{fig:average_power}, respectively. In the bottom panel, we plot the offset of the maximally dense cell from the center of mass of the halo, $R_{\rm sh}$, as a function of time. The offset is less than or of the order of the soliton radius, its maximum value being $\sim 2.3\kpc$. This random-walk of the soliton has been previously discussed in \cite{schive20} and \cite{li20}.

\section{Random Motion of Nuclear Objects} 
\label{sec:particles}

At any time, the net force acting on an object orbiting inside a virialized FDM halo consists of a time-invariant, smooth force derived from the time-and-shell averaged potential, plus a time-variable, stochastic force due to the perturbations in the potential about its time-and-shell averaged value. While the smooth force conserves energy, the stochastic force results in a diffusive heating of the object, causing it to increase its binding energy and move outwards to larger and larger radii (in an orbit-averaged sense). In addition, the motion of the object creates a wake behind it that induces a friction force proportional to its mass \citep[e.g.,][]{lancaster20}. Known as dynamical friction, this retarding force transfers energy from the object to the FDM halo, thereby opposing the diffusive heating. As a result, the object's orbital energy continues to evolve until an equilibrium is established in which the diffusive heating rate is balanced by the frictional cooling rate.

In this section, we investigate what this implies for (baryonic) objects that one typically expects to find located at (or near) the very center of a galaxy potential well (i.e., SMBHs and nuclear star clusters). We examine this using numerical simulations of the FDM halo described in Section~\ref{sec:isolated_fdm_halo} in which we position a massive object (modeled as a point particle) at rest at the center of the soliton, at some random time, and study its subsequent evolution. Because of the random nature of the motion and density oscillations of the soliton, we can study the realization-to-realization variance in the motion of the object by positioning it at rest at the soliton center at different times in one and the same FDM halo, rather than having to resort to different halos. We use this method to simulate 30 realizations each for particle masses of $m=10^{5}$, $10^{6}$, $5 \times 10^{6}$, and $10^{7} M_{\odot}$. In each case, the particle is positioned at rest at the densest cell of the soliton at a different time of introduction, $t_0$, and the system is then evolved for $20 \Gyr$. 

\begin{figure*}
    \centering
    \includegraphics[width=0.9\textwidth]{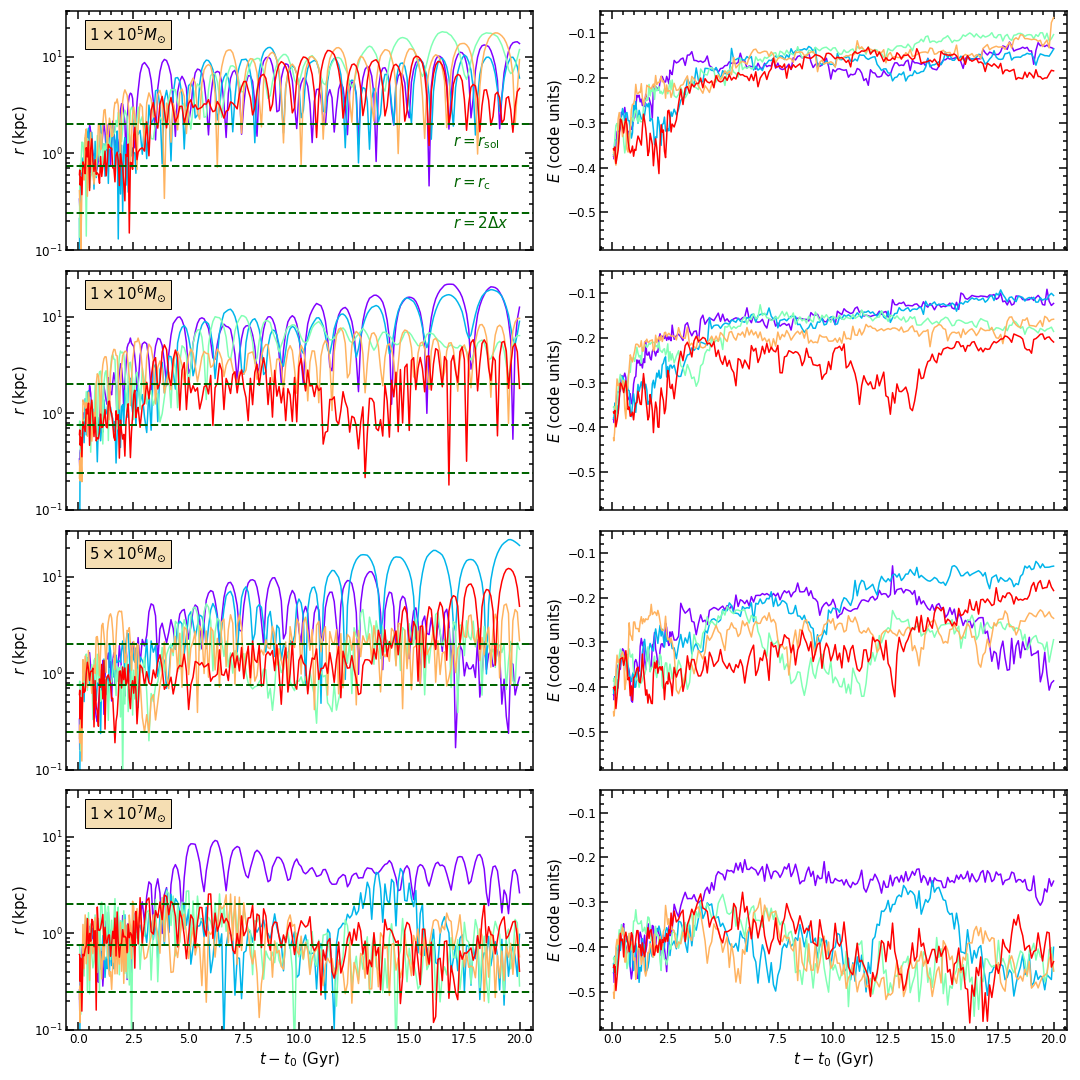}
    \caption{The different curves in the left and right-hand panels of each row show the evolution of the 3D position and specific energy, both measured in the soliton frame, of a particle of mass, $m$ placed at rest at the soliton center at five different instants of time, $t_0$. From top to bottom, the different rows correspond to $m = 10^{5}$, $10^{6}$, $5 \times 10^{6}$, and $10^{7} M_{\odot}$, respectively. The realization-to-realization variance in the evolution of the particle is evident, depending on the exact interplay between the three different forces acting on it - that from the time-and-shell averaged potential, that from the fluctuations in the potential about its time-and-shell averaged value, and that from dynamical friction. The dashed, green lines in the left-hand panels indicate the time-averaged soliton radius, $r_{\rm sol}$, core-radius, $r_\rmc$, and the spatial resolution of the simulations, $2 \Delta x$, as labeled.}
    \label{fig:sample_realizations}
\end{figure*}

\subsection{A Specific Example}
\label{sec:example}

As an example, the four top panels of Figure~\ref{fig:test_particle_motion} show snapshots of one of these simulations with a particle mass of $m=10^5 \Msun$. Each panel shows the projected density in a region of  $10 \kpc \times 10 \kpc$ centered on the soliton at $t=t_{0}$. The white dots indicate the instantaneous positions of the particle, while the white curves in the top-right, bottom-left, and bottom-right panels show its orbital evolution (with respect to the position of the soliton at $t=t_0$) during the preceding periods from $0-2\Gyr$, $2-5\Gyr$, and $5-10 \Gyr$, respectively. At early times, the particle is more or less confined to the soliton, where it undergoes a random motion, driven by the stochastic force arising from the potential fluctuations within the soliton. At later times, the particle has diffused out of the soliton and moves around on what resembles a typical rosette orbit. However, due to the potential fluctuations within the halo, the orbit is continuously undergoing weak perturbations that continue to diffuse the particle out to larger and larger radii.

This is more apparent from the bottom panel, which shows the orbital evolution of the particle with respect to both the instantaneous soliton center (green curve) and the halo center of mass (orange curve), respectively. For comparison, the soliton's motion with respect to the halo center of mass is indicated by the blue curve, while the dashed, horizontal line indicates the time-averaged soliton radius. During the first $\sim 5 \Gyr$ the particle remains largely confined to the wobbling soliton while slowly diffusing outwards. At later times, its orbital energy has increased such that it remains outside of the soliton while its apo- and peri-centric distances continue to increase due to ongoing heating caused by gravitational interactions with the soliton and the halo granules (i.e., quasiparticles). 

\subsection{Summary Statistics}
\label{sec:statistics}

Figure~\ref{fig:sample_realizations} illustrates the realization-to-realization variance in the particle motion. Each row shows the evolution in the distance from the soliton center, $r$ (left-hand panels), and energy, $E$ (right-hand panels), of the particle for a random subset of 5 of the 30 realizations. Here, $E=0.5 v^{2} + \phi$, where $v$ is the particle's velocity with respect to the instantaneous velocity of the soliton, and $\phi$ is the gravitational potential at its location. From top to bottom, the different rows correspond to particles of mass $m=10^{5}$, $10^{6}$, $5 \times 10^{6}$, and $10^{7} M_{\odot}$, respectively. The dashed, horizontal, green lines in the left-hand panels indicate the spatial resolution of the simulations ($2 \Delta x$), the time-averaged core radius, and the soliton radius, as labeled.

As expected from a stochastic force field, there is significant realization-to-realization variance in the orbital evolution of the particle. We, therefore, combine the results from 30 realizations for each mass, $m$, to statistically characterize their evolution. For comparison with the massive particles, we also evolve $10^{4}$ massless particles, placed simultaneously at rest at random positions within the core radius of the soliton. As massless particles do not interact with each other, we can evolve any number of them simultaneously. In particular, the (late-time) trend in the ensemble-averaged properties of the massless particles should be the same as if they were evolved individually by placing them at rest at the soliton center at different times. 

\begin{figure*}
    \centering
    \includegraphics[width=0.9\textwidth]{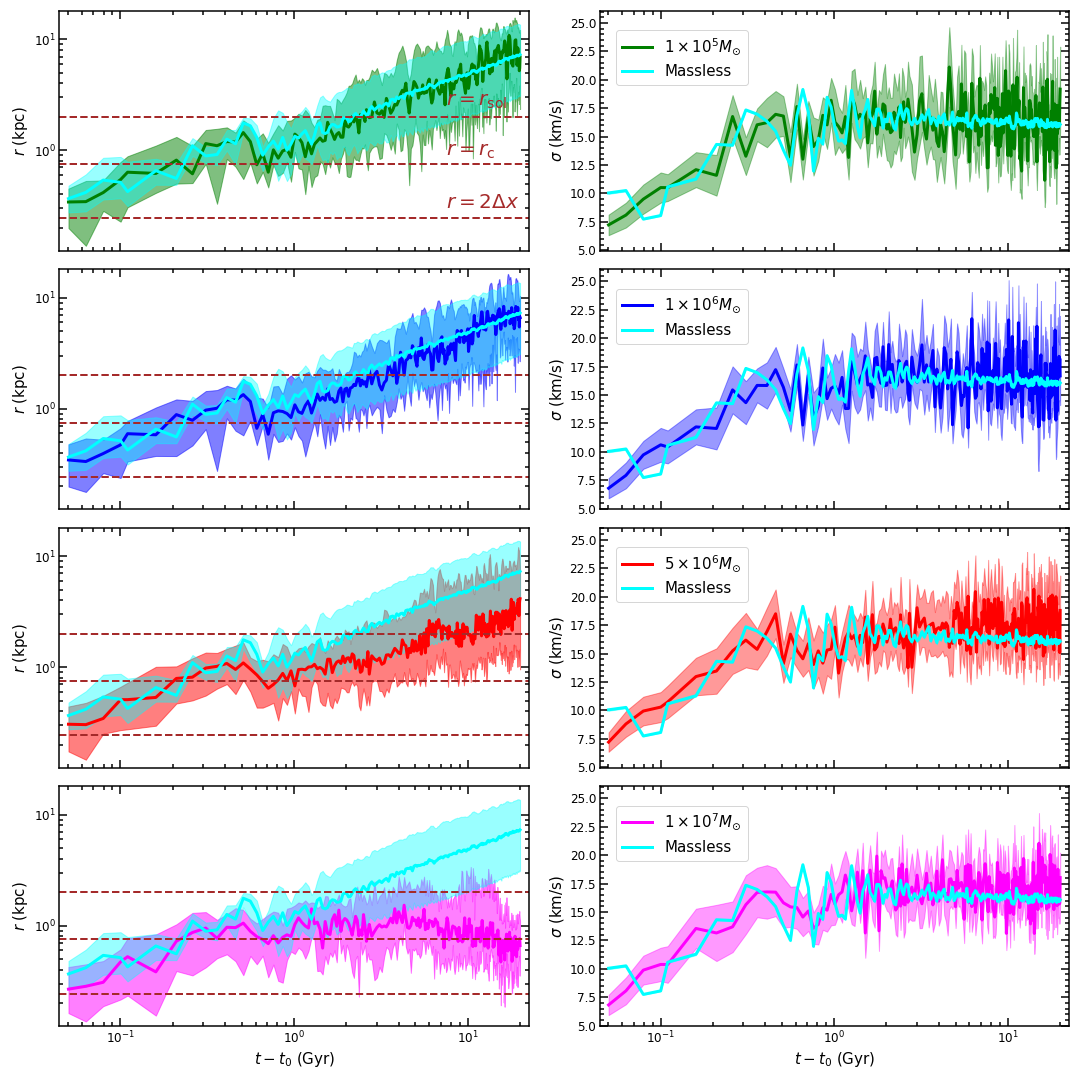}
    \caption{The left- and right-hand panels show the evolution of the median radius, $r_{\rm med}$, along with the $16^{\rm th}-84^{\rm th}$ percentile variation in $r$, and the velocity dispersion, $\sigma$, along with its $95 \%$ confidence interval, for an ensemble of particles of mass, $m$, both measured in the soliton frame. From top to bottom, the different rows correspond to $m=10^{5}$, $10^{6}$, $5 \times 10^{6}$, and $10^{7} \Msun$, respectively. The results for massless particles are overplotted in cyan in each panel. The dashed, horizontal, lines indicate the time-averaged soliton radius, $r_{\rm sol}$, core-radius, $r_\rmc$, and the spatial resolution of the simulations, $2 \Delta x$, as labeled. The outward diffusion for $m \leq 10^{6} \Msun$ is almost identical to that of the massless particles, indicating that, on average, they experience negligible dynamical friction within the runtime of the simulations. In contrast, for $m \geq 5 \times 10^{6} \Msun$, dynamical friction is important, leading to a reduced outward diffusion (and stalling for $m=10^{7} \Msun$). See text for details.}
    \label{fig:rms_radius_dispersion}
\end{figure*}

From top to bottom, the solid curves in the left-hand panels of Figure~\ref{fig:rms_radius_dispersion} show the evolution of the median distance from the soliton center, $r_{\rm med}$, calculated from the 30 realizations, for $m=10^{5}$, $10^{6}$, $5 \times 10^{6}$, and $10^{7} M_{\odot}$, respectively, over $20 \Gyr$. In each case, the associated envelope highlights the $16^{\rm th}-84^{\rm th}$ percentile variation in $r$. The solid curves in the right-hand panels show the $1D$ velocity dispersion, $\sigma$, of the particle ensemble, defined as $\sigma=\sqrt{\langle v^{2} \rangle/3}$, where $v$ is the velocity of the particle with respect to the soliton. The associated envelopes indicate the $95 \%$ confidence intervals in $\sigma$, calculated using the Jackknife method. In each panel, the relevant statistic for the $10^{4}$ massless particles is shown in cyan. Because of the large number of realizations, the Jackknife error in $\sigma$ of the massless particles is very small, rendering the $95\%$ confidence interval narrower than or comparable to the line width. The dashed, horizontal, brown lines in the left-hand panels indicate the spatial resolution of the simulations ($2 \Delta x$), the time-averaged core radius, and the soliton radius, as labeled. 

With no dynamical friction acting on them to counteract the heating effect of FDM potential fluctuations, the median radius of the massless particles keeps increasing with time. In contrast, $r_{\rm med}$ of massive particles is expected to increase until the ensemble-averaged heating rate from FDM potential fluctuations balances the ensemble-averaged cooling rate due to dynamical friction, thereby reaching an equilibrium state in which $r_{\rm med}$ and $\sigma$ become stationary. 

For $m=10^{5}$ and $10^{6} M_{\odot}$, the evolution of $r_{\rm med}$ and $\sigma$ is virtually indistinguishable from that of the massless particles, indicating that their cooling rate remains small compared to their heating rate over the entire runtime of the simulations. Consequently, even after $20\Gyr$, these masses have not reached equilibrium. For $m=5 \times 10^6$ and $10^{7} M_{\odot}$, the evolution of $r_{\rm med}$ is similar to that of the massless particles only up to $\sim 0.4 \Gyr$. After that, dynamical friction becomes important, which is stronger for the more massive particle, and the subsequent evolution of $r_{\rm med}$ is different from that of the massless particles. For $m=5 \times 10^6 M_{\odot}$, $r_{\rm med}$ keeps increasing, and the particles do not reach equilibrium within the runtime of our simulations. However, on average, it clearly diffuses less rapidly than a massless particle, reaching a median radius of $\sim 3\kpc$ after $20 \Gyr$, compared to $\sim 7\kpc$ in the massless case.  For $m=10^{7} M_{\odot}$, the median radius seems to stall around $1 \kpc$ after only $\sim 1 \Gyr$, indicating a rough balance between the ensemble-averaged heating and cooling rates. 

Note that the balance between heating and cooling for $m=10^{7} M_{\odot}$ is not perfect and that there is a weak decreasing trend in $r_{\rm med}$ at late times ($t-t_{0} \gtrsim 6 \Gyr$). We find that this decay is well described by an exponential with a decay time of $\sim 30 \Gyr$ and suspect that it is due to the slow but gradual increase in the mass of the soliton over time, as a result of the gravitational cooling process discussed in Section~\ref{sec:Intro} \citep[see also][]{hui17}. This increased inertia reduces the overall amplitude of the soliton random walk and the heating associated with it.  As a consequence, the radius where heating balances cooling is a quasi-equilibrium, which shrinks secularly. Note, though, that the relatively long characteristic timescale of this phenomenon implies that it is of little astrophysical significance.

The evolution of $\sigma$ reveals remarkably little mass dependence. In each case, it starts out with a phase of $\sim 0.4 \Gyr$ in which the velocity dispersion increases rapidly from zero (each particle initially starts out with zero velocity with respect to the soliton) to about $17 \kms$, after which it changes only very slowly. During this later phase, the evolution of $\sigma$ for $m=10^{5}$ and $10^{6} M_{\odot}$ remains almost identical to that of the massless particles. For $m=5 \times 10^{6}$ and $10^{7} M_{\odot}$, $\sigma$ reaches values that are only mildly higher than that of the massless particles. 

\section{Comparison with Kinetic Theory}
\label{sec:discusssion}

The simulations discussed above are limited to a single halo mass ($\Mvir = 6.6\times 10^9 \Msun$) and a single boson mass ($m_\rmb = 8 \times 10^{-23} \eV$). In this section, we compare our simulation results with theoretical predictions based on kinetic theory. The goal is to devise a theoretical underpinning for the outward diffusion of nuclear objects that can be used to make predictions for any halo mass and any boson mass, without having to resort to CPU intensive simulations.

As noted in Section~\ref{sec:velocity_dispersion_profile}, outside the soliton, an FDM halo can be modeled as a system of quasiparticles. The effective mass, $m_{\rm eff}$, of these quasiparticles is given by Equation~\ref{effective_mass}, and their velocity dispersion is the same as $\sigma_{\rmh}$, which is approximately equal to $\sigma_{\rm Jeans}/\sqrt{2}$ (see Figure~\ref{fig:halo_dispersion}). The evolution of the energy of a particle of mass, $m$ as it moves through this sea of quasiparticles can be described in terms of the specific energy diffusion coefficient, $D [\Delta E]$, which is given by
\begin{equation}
D [\Delta E]= \frac{1}{2}D[(\Delta v_{||})^2]+ \frac{1}{2}D[(\Delta v_{\perp})^2] +v D[\Delta v_{||}] \, .
\label{energy_diffusion}
\end{equation}
Here, $v$ is the velocity of the particle, $D[(\Delta v_{||})]$ is the first-order velocity diffusion coefficient parallel to $v$, also known as the dynamical friction coefficient, and $D[(\Delta v_{||})^2]$ and $D[(\Delta v_{\perp})^2]$ are the second-order velocity diffusion coefficients parallel and perpendicular to $v$, respectively. 

Assuming a Gaussian velocity distribution for the quasiparticles in the FDM halo, \citet[]{bar-or19}, \citet{chavanis20}, and \citet{elzant20a} show that the velocity diffusion coefficients are given by
\begin{align}
v \, D[\Delta v_{||}] & = -\calD \, X_{\rm eff} \left[\mathbb{G}(X_{\rm eff}) + \mu_{\rm eff} \mathbb{G}(X)\right] \,, \label{fo_para} \\
D[(\Delta v_{||})^2] &= \calD \, \frac{\mathbb{G}(X_{\rm eff})}{X_{\rm eff}} \,, \label{so_para} \\
D[(\Delta v_{\perp})^2] &= \calD \, \frac{{\rm erf}(X_{\rm eff})-\mathbb{G}(X_{\rm eff})}{X_{\rm eff}} \,, \label{so_perp}
\end{align}
with
\begin{equation}
\calD = \frac{4 \sqrt{2} \pi G^2 \rho \, m_{\rm eff}}{\sigma_{\rmh}} \, \ln\Lambda_{\rm FDM} \,.
\label{CalD}
\end{equation}
Here, $\rho$ and $\sigma_\rmh$ are the time-and-shell averaged density and velocity dispersion, respectively, $\ln\Lambda_{\rm FDM}$ is the Coulomb logarithm, given by
\begin{equation}
\ln\Lambda_{\rm FDM}=\ln \left( \frac{4 \pi r}{\lambda_{\rm db}} \right) \,,  
\end{equation}
where $r$ is the distance from the soliton center,
\begin{equation}
\mathbb{G}(X) = \frac{1}{2X^2} \left[{\rm erf}(X) - \frac{2X}{\sqrt{\pi}} \rme^{-X^2} \right] \, ,
\label{G}
\end{equation}
$X_{\rm eff} = v/\sqrt{2} \sigma_{\rmh}$, $X = v/\sqrt{2} \sigma_{\rm Jeans}$, and 
\begin{equation}
\mu_{\rm eff}=\frac{m}{m_{\rm eff}} \frac{\sigma_{\rm h}^2}{ \sigma_{\rm Jeans}^2} \, .
\label{mu_eff}
\end{equation}
Note that the second term in Equation~\ref{fo_para} is directly proportional to the mass of the particle, indicating that cooling due to dynamical friction becomes dominant over diffusive heating for sufficiently massive particles.

Although the above diffusion coefficients are derived assuming an infinite, homogeneous sea of quasiparticles, in which the unperturbed trajectory of the object is a straight line, it is common practice to assume that they hold locally. Using this local approximation, it is possible to investigate the orbital evolution of an object in our FDM halo using the following Monte-Carlo method. At $t=0$, we uniformly distribute $400$ equal mass particles within a sphere of diameter, $\Delta x=0.122 \kpc$ centered at $r=0$, each with zero velocity. Each particle is then evolved for $200 \Gyr$ in the smooth, time-and-shell averaged potential of our FDM halo, using a $4^{\rm th}$-order Runge-Kutta integrator with a time step of $\Delta t = 10^{5} \yr$.\footnote{We have verified that our results do not change if we use smaller time steps.} At the end of each time step, though, the velocity, {\bf v}, of each particle is perturbed by a distinct $\Delta {\bf v}$, the components of which are drawn from a Gaussian distribution with mean $\mu_{i}=D[\Delta v_{i}] \Delta t$ and covariance matrix $C_{ij}=D[\Delta v_{i} \Delta v_{j}] \Delta t$. $D[\Delta v_{i}]$ and $D[\Delta v_{i} \Delta v_{j}]$ are the first and second-order velocity diffusion coefficients along the three Cartesian axes, which are given by 
\begin{align}
D[\Delta v_{i}] &= \frac{v_{i}}{v} D[\Delta v_{||}] \, , \\
D[\Delta v_{i} \Delta v_{j}] &= \frac{v_{i}v_{j}}{v^2} \big\{ D[(\Delta v_{||})^2]-\frac{1}{2} D[(\Delta v_{\perp})^2] \big\} \nonumber \\ 
&+\frac{1}{2} \delta_{ij} D[(\Delta v_{\perp})^2] \, ,
\end{align}
\citep[e.g.,][]{binney08}. This models both the diffusive heating as well as the cooling due to dynamical friction. 

\begin{figure*}
    \centering
    \includegraphics[width=1.\textwidth]{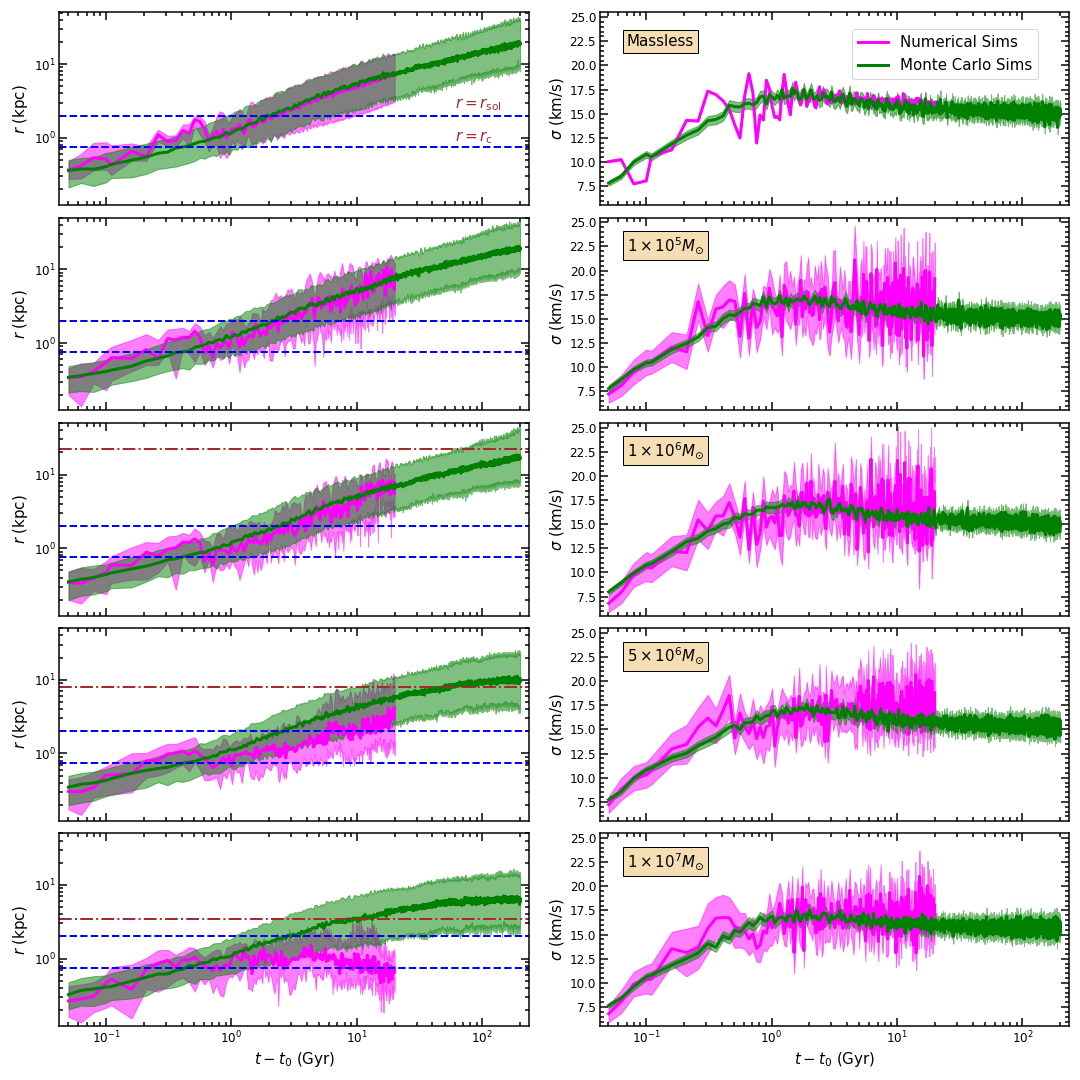}
    \caption{Same as Figure~\ref{fig:rms_radius_dispersion}, except that here the results from the numerical simulations (in magenta) are overplotted on those obtained using the Monte Carlo method (in green) described in Section~\ref{sec:discusssion}. Different rows correspond to a different a particle mass, $m$, as indicated. For $m \leq 10^{6} \Msun$, the results from the numerical and Monte-Carlo simulations are in good agreement. For $m=5 \geq 10^{6} \Msun$ though, the Monte-Carlo results significantly overpredict the outward diffusion compared to the numerical simulations. The horizontal, brown, dot-dashed lines mark the radius where $\mu_{\rm eff}$, defined according to Equation~\ref{mu_eff}, is equal to $1$, which is off-scale for $m=10^{5} \Msun$ and not defined for massless particles. The horizontal, blue, dashed lines indicate the time-averaged core radius, $r_\rmc$, and soliton radius, $r_{\rm sol}$, as labeled. See text for details.}
    \label{fig:compare_sims_mc}
\end{figure*}

However, there is one caveat. Strictly speaking, the velocity diffusion coefficients, given by Equations~\ref{fo_para}--\ref{so_perp}, are only valid well outside of the soliton. After all, they are obtained under the assumption that the diffusion is dominated by many weak encounters (with quasiparticles) rather than by strong encounters. This assumption is no longer valid near the soliton, which acts as a single quasiparticle that dominates the perturbations experienced by the particle. In fact, because of the strong increase in $\rho$ and decrease in $\sigma_\rmh$ as one approaches the center of the solitonic core, the diffusion coefficients, which scale as $\rho^2/\sigma^4_\rmh$, become excessively large. We therefore construct an `effective model' in which we replace the values of the diffusion coefficients within some specific cut-off radius, $\rcut$, with their values {\it at} $\rcut$. After some trial and error, we find that $\rcut = 2.3 r_\rmc \simeq 0.85\, r_{\rm sol}$ yields results that are in good agreement with the simulation results, at least for low mass particles (see below). Hence, all Monte-Carlo results shown below have adopted this effective model with this particular value for $\rcut$.

Figure~\ref{fig:compare_sims_mc}, which is similar to Figure~\ref{fig:rms_radius_dispersion}, compares the results from these Monte-Carlo simulations (in green) to our numerical simulations (in magenta). Different rows correspond to different particle masses, as indicated. Note that the Monte-Carlo results for $m \leq 10^6 \Msun \simeq 0.003 M_{\rm sol}$ (top three rows) are in good agreement with the numerical simulations. This indicates that the diffusive heating due to interactions with both the soliton and the quasiparticles can be effectively modeled by the diffusion coefficients of \citet[][]{bar-or19}, \citet{chavanis20}, and \citet[][]{elzant20a} by simply using a cut-off radius, $\rcut \simeq 2.3 r_\rmc$, as described above. In fact, this effective model can even be used to adequately describe the outward diffusion of low mass objects ($m \leq 10^{6} M_{\odot}$) within the confines of the soliton itself.

In agreement with the numerical simulations, for all three particle masses, the Monte-Carlo predictions are virtually indistinguishable, indicating that dynamical friction remains insignificant, i.e., heating continues to dominate over cooling, at least for the full $200 \Gyr$ of evolution probed by the Monte Carlo simulations. In the case of the more massive particles with  $m=5 \times 10^6$ and $10^{7} \Msun$ (i.e., $m \gta 0.01 M_{\rm sol}$), dynamical friction can no longer be neglected, resulting in a reduced diffusion rate outwards. In general, though, the Monte Carlo results significantly overpredict the outward diffusion compared to the numerical simulations. Apparently, whereas our effective treatment seems adequate to describe the diffusion of low-mass particles in or near the solitonic core, it significantly underestimates the dynamical cooling impacting more massive objects.

\subsection{Stalling Radius}
\label{sec:stalling}

The outward diffusion of particles of a given mass continues until the ensemble-averaged heating and cooling rates become equal, and the particle ensemble attains equilibrium with the halo. Using Monte-Carlo simulations similar in nature to those presented here, \citet[][]{bar-or19} investigated the inspiral of a $4 \times 10^{5} \Msun$ particle in an isothermal FDM halo (without a soliton) with a boson mass of $m_\rmb = 10^{-21} \rm eV$ and found that when such as equilibrium is attained, the median particle radius settles around a value of $r_{\rm stall}$, defined as the radius where $\mu_{\rm eff}=1$. Using this criterion, and the time-and-shell averaged density and velocity dispersion profiles of our FDM halo, we infer that $r_{\rm stall}$ should decrease from $66 \kpc$ (which lies beyond the halo's virial radius) for $m=10^5 \Msun$ to $3.4\kpc$ for $m=10^7 \Msun$.

The dot-dashed, horizontal, brown lines in the left-hand panels of Figure~\ref{fig:compare_sims_mc} mark these predicted stalling radii for our simulations. For $m = 10^5 \Msun$ (second row from the top), no brown line is visible, indicating that the $r_{\rm stall}$ value is off scale. In the case of $m = 5 \times 10^6 \Msun$, equilibration is predicted at $r_{\rm stall} \simeq 8 \kpc$. This is somewhat lower than the asymptotic value of $r_{\rm med}$ for the ensemble in the Monte-Carlo simulations, which is roughly $10 \kpc$ (corresponding to $\mu_{\rm eff} \simeq 1.3$). Finally, for $m=10^7 \Msun$, the predicted $r_{\rm stall}$ ($\sim 3.4\kpc$) is a poor match to the Monte-Carlo simulation results, which suggest that stalling happens at $r_{\rm stall} \sim 6 \kpc$, corresponding to $\mu_{\rm eff} \simeq 1.6$). 

Hence, we conclude that $\mu_{\rm eff}=1$ is a rather poor predictor for the stalling radius. Apparently, there is no single value for $\mu_{\rm eff}$ that can be used to estimate the median radius at which an ensemble of objects of a given mass is expected to equilibrate. Instead, the value of $\mu_{\rm eff}$ corresponding to $r_{\rm stall}$ seems to depend on the particle mass, $m$, and also possibly on the detailed potential of the host system. In principle, one could try to compute $r_{\rm stall}$ by balancing the ensemble-averaged heating and cooling rates, but this requires integrating the diffusion coefficients over phase-space, weighted by the (time-dependent) distribution function of the ensemble, which is unknown. Hence, predicting a stalling radius based on the kinetic theory of weak encounters is best done using Monte-Carlo simulations, rather than relying on a simple $\mu_{\rm eff}$ criterion. 

We caution, though, that the effective treatment of diffusion in or near the solitonic core advocated here is only reliable for relatively low mass particles ($m \lta 0.003 M_{\rm sol}$). For more massive objects the Monte-Carlo simulations clearly fail to  reproduce the behavior seen in the numerical simulations. In particular, for $m=10^7\Msun$, stalling actually happens at $r_{\rm stall} \sim 1 \kpc$ after only $\sim 1 \Gyr$ in clear disagreement with the Monte-Carlo simulations, which predict stalling at around $6 \kpc$ after $\sim 60\Gyr$.

\section{Summary \& Conclusions}
\label{sec:concl}

Fuzzy dark matter is assumed to be made up of ultralight bosonic particles with a mass $m_\rmb \sim 10^{-22} \eV$. Simulations of structure formation in an FDM universe have shown that dark matter halos consist of a central soliton (representing the ground-state of the SP equation) surrounded by an NFW-like envelope made up of excited states. Because of wave-interference, the halo is characterized by order unity density fluctuations (wave-granularity), which can be treated as quasiparticles with an effective mass, $m_{\rm eff} \propto \rho \lambda^3_{\rm db}$, that are moving through the halo with a velocity dispersion, $\sigma_\rmh \simeq \sigma_{\rm Jeans}/\sqrt{2}$. 

Due to interference between the ground state (the soliton) and the excited states that make up the halo surrounding the soliton, the latter is subjected to stochastic forces that cause it to undergo a random walk within the central confines of the halo. In addition, the soliton experiences order unity temporal oscillations in its density. In this paper, we have used numerical and Monte-Carlo simulations to investigate these effects and their impact on nuclear objects such as SMBHs or (dense) star clusters. 

Using high-resolution ($\Delta x = 122\pc$) simulations of a $M \simeq 6.6\times 10^9\Msun$ FDM halo, extracted from a larger cosmological volume, with $m_\rmb=8 \times 10^{-23} \eV$, we first studied the oscillations and the random walk of the soliton. In agreement with \cite{veltmaat18}, the density of the soliton fluctuates with a characteristic frequency of $f \sim 4 \Gyr^{-1}$ (see equation~\ref{oscillation_freq}). In addition, it undergoes a confined random walk, resulting in an offset from the center of mass of the halo that is of order its own radius \citep[see also][]{schive20, li20}.

Outside of the soliton, the power spectrum of density fluctuations peaks approximately at the mode corresponding to $0.35 \lambda_{\rm db}$, consistent with the expected size of the quasiparticles \citep[see][]{bar-or19,chavanis20,elzant20a}, and the velocity dispersion associated with the fluctuations (i.e., the velocity dispersion of the quasi-particles) provides roughly half of the pressure support against gravity, the other half coming from quantum pressure \citep[see also][]{mocz17}.

A naked soliton (i.e., without a surrounding halo) has a time-invariant, spherically symmetric density profile (and consequently potential). Therefore, a baryonic object placed at rest at the center of a naked soliton would feel no net force with respect to the soliton and remain at rest forever. However, in an FDM halo, the soliton interacts with the halo envelope, and while it is subject to both gravity and gradients in quantum pressure, the cluster only feels the former. Hence, the two respond differently, initiating a random walk of the cluster with respect to the soliton, in which the cluster's orbit continues to be gravitationally perturbed by the wobbling, oscillating soliton.

In order to study the mass-dependence of this random motion, we placed point particles of mass $m$ at rest (at some random time) at the center of the soliton, and evolved them for a duration of $20 \Gyr$. Our key results are as follows:
\begin{itemize}
  \item Nuclear objects are unable to remain at the center of the soliton. On average, such objects with mass $m \lta 0.003 M_{\rm sol}$ diffuse out of the soliton in $\sim 3 \Gyr$, while those with $m \gta 0.01 M_{\rm sol}$ experience strong dynamical friction such that they never diffuse much beyond the soliton radius.
  \item Nuclear objects with mass $m \lta 0.003 M_{\rm sol}$ diffuse in the same way as massless particles, indicating that they experience negligible dynamical friction.
  \item The equilibration time for nuclear objects exceeds the Hubble time for $m \lta 0.01 M_{\rm sol}$. Hence, their average displacement from the center of the soliton depends strongly on their age.
\end{itemize}

We compared these results to Monte-Carlo simulations in which the trajectories of particles of a given mass were integrated in the smooth, time-and-shell averaged halo potential and subjected to velocity kicks drawn from the diffusion coefficients derived by \cite{bar-or19}, \citet{chavanis20}, and \cite{elzant20a}. Near the soliton, the diffusion is dominated by gravitational interactions with the wobbling soliton, which is not adequately described by the kinetic theory of weak encounters that underlies the derivation of these diffusion coefficients. We, therefore, opted for an `effective' treatment, in which the diffusion coefficients inside a cut-off radius, $\rcut \simeq 2.3 r_\rmc \simeq 0.85 r_{\rm sol}$, are replaced by their values at $\rcut$. The resulting Monte-Carlo simulations can accurately reproduce the numerical simulation results for nuclear objects with mass less than approximately 0.3 percent of the soliton mass. For more massive objects ($\gta 0.01 M_{\rm sol}$), though, the Monte-Carlo simulations underestimate the amount of dynamical friction, rendering the semi-analytical effective treatment based on the diffusion coefficients unreliable.

Our main conclusion is that FDM models predict that massive objects that one typically expects to find at the centers of galaxies, such as a SMBHs and nuclear star clusters, should be offset (in a statistical sense) from the center of mass of the stellar body. This is a clear, testable prediction that has the potential to validate the FDM picture or at least to put constraints on $m_\rmb$. But, the specific results presented in this paper are only valid for an isolated halo of of mass $M_{\rm vir} = 6.6 \times 10^9 \Msun$ in a FDM model with $m_\rmb=8 \times 10^{-23} \eV$. To put this in perspective,  we extrapolate the average stellar mass-halo mass relation at $z \sim 0$ to the low mass end to infer that the expected stellar mass for a galaxy in our halo is $M_{\ast} \approx 0.2 - 2 \times 10^6 \Msun$ \citep[e.g.,][]{yang09, moster10}. This is the typical stellar mass for classical Milky Way dwarf spheroidals such as Draco or Sculptor \citep[see e.g.,][]{McConnachie12}. Hence, any SMBH or nuclear cluster in the center of such a galaxy is likely to have a mass $m \ll 10^6 \Msun$. As shown in this paper, for $m_\rmb=8 \times 10^{-23} \eV$, such objects have equilibration times much larger than the Hubble time and behave like massless particles for all practical purposes. Therefore, their offsets from the galaxy center will depend on their age.

However, there are several caveats to keep in mind. First of all, the result presented here only applies to {\it isolated} FDM halos. Once an FDM halo becomes a subhalo, its outer envelope will be tidally stripped, which can drastically reduce the wave granularity and hence the wobbling of the soliton \citep[see][]{schive20}. In the extreme case only a naked soliton survives, and the fluctuations cease entirely, after which any massive object that has not yet been tidally stripped will slowly sink back to the center of the soliton due to dynamical friction. Hence, in the case of Draco or Sculptor, the predicted offsets will also depend on how long ago their host halos became subhalos of the Milky Way, and how much of their mass has since been stripped. Secondly, the offsets are likely to depend significantly on both halo and boson mass. Although we anticipate that our effective treatment can also be used to model the outward diffusion of nuclear objects in halos of different mass and for different $m_\rmb$, this requires careful testing using numerical simulations such as those presented here. This will be the topic of a follow-up paper, in which we will also aim to quantify the expected offsets of SMBHs and dense nuclear star clusters as a function of both halo and boson mass.

\acknowledgments

The authors thank the anonymous referee for insightful feedback that helped in improving the manuscript. DDC is grateful to Uddipan Banik, Sheridan Green, and Puskar Mondal for valuable discussions and thanks the Yale Center for Research Computing for guidance and use of the research computing infrastructure, specifically the Grace cluster. FvdB is supported by the National Aeronautics and Space Administration through Grant Nos. 17-ATP17-0028 and 19-ATP19-0059 issued as part of the Astrophysics Theory Program and received additional support from the Klaus Tschira foundation. V.H.R. acknowledges support by YCAA Prize postdoctoral fellowship.  H. S. acknowledges funding support from the Jade Mountain Young Scholar Award No. NTU-109V0201, sponsored by the Ministry of Education, Taiwan. This research is partially supported by the Ministry of Science and Technology (MOST) of Taiwan under Grants No. MOST 107-2119-M-002-036-MY3 and No. MOST 108-2112-M-002-023-MY3, and the NTU Core Consortium project under Grants No. NTU-CC-108L893401 and No. NTU-CC-108L893402.

\bibliography{ms}{}
\bibliographystyle{aasjournal}

\end{document}